\documentstyle[12pt]{article}
\def\bfg #1{{\mbox{\boldmath $#1$}}}
\large
\begin{document}
\begin{center}
 {\Large \bf  Spin observables of the reaction
 $pd\to~^3He\eta$ and quasi-bound $^3He-\eta$ pole}
 \end{center}
  \centerline{  Yu.N.~Uzikov\footnote{
{\it e-mail address}: uzikov@nusun.jinr.ru}}
\centerline{ Joint Institute for Nuclear Research, Dubna, Russia}
\begin{abstract}
  A formalism for spin observables of the reaction
 $pd\to ~^3He\eta$ is derived in a model independent way.
 The general case with a full set of six independent spin 
 amplitudes is studied. Furthermore, approximations  by five and  
 four spin amplitudes
 are investigated in the near threshold region.
 This region is of great interest to search
 for a quasi-bound  $^3He-\eta$ state, in particular, by  measurement
 of energy dependence of  relative  phases of  s- and
 p-wave amplitudes. 
 Complete polarization experiments, allowing determination of 
 spin amplitudes, are analyzed. It is shown 
 that measurement of only analyzing powers and spin correlation
 coefficients hardly allows  one to separate the
 s- and p-wave amplitudes, but additional  measurement
 of polarization  transfer coefficients simplifies this problem.
 Specific observables, given by products of one s-
 and one p-wave amplitudes, are found.  Measurement of these
 observables will provide new independent information
 on the $^3He-\eta$ pole position.

\end{abstract}


\vspace{1cm}

\section{Introduction}

The reaction $pd\to~^3He\eta$ has been arousing interest since the
 first measurement performed at SATURNE \cite{berger}.
 Strong enhancement of the cross section
 of this reaction was found
 near the threshold and confirmed later on in Ref. \cite{mayer}.
 This enhancement  was interpreted as  a strong final state
  interaction  effect arising due to  presence of a 
 quasi-bound $^3He-\eta$ state \cite{cwilkin93}. This
 idea was supported by the calculation of this reaction within
 a two-step model 
\cite{KLU,WF} and  parameters of the quasi-bound state were
 estimated \cite{KLU} from  fit to the data.
 An independent experimental
indication of existence of the
quasi-bound $^3He-\eta$ state was found in measurement of
 the photo-production of the $\eta-$ meson on $~^3He$
in the reaction $\gamma ~^3He\to ~^3He\eta$
\cite{photo}, although this interpretation
was questioned in Ref. \cite{hanhart}.
 Microscopic four-body calculations support
 quasi-bound \cite{rakit} or anti-bound (virtual)
 \cite{fix} states in the $^3He-\eta$ system within the existing
uncertainties for the elementary $\eta N$ interaction.
A special type of experiments
is required to decide whether the $^3He-\eta$ system,
if it really exists, 
is in a  quasi-bound (for strong enough $\eta N$ interaction)
or anti-bound  (for rather weak $\eta N$ interaction) state \cite{baru}.

 Recently the near threshold cross section
 of the $pd\to~^3He\eta$ reaction  was measured at COSY
\cite{mersmann,smyrski}
 with  a high precision.
 According to Ref.\cite{mersmann}, the pole
 corresponding to the quasi-bound (or anti-bound) state
 is located on the excitation
 energy plane at the point
$Q_0=[(-0.30\pm0.15_{stat}\pm0.04_{syst})\pm
\imath(0.21\pm0.29_{stat}\pm0.06_{syst}]$ MeV, that is
 very close to the threshold.
 As was noted in \cite{cwilkin07}, the presence
 of the pole in the s-wave amplitude of this reaction
 must lead to
 fast variation not only in its magnitude, but also
 in the phase. The latter provides a new criterion for identification of
 the pole.
 The specific phase behaviour comes
 from the following analytical form of the
 s-wave  amplitude near the pole:
\begin{equation}
\label{pole}
f(p_\eta)=\frac{\xi}{p_\eta-ip_0},
\end{equation}
where $p_\eta$ is the (real) c.m.s.  momentum
of the $\eta$-meson, $p_0$ is
the (complex) pole point directly  related to the
 energy $Q_0$, and
  $\xi$ is a smooth function of  $p_\eta$.
On the other hand, p-waves are
 expected to exhibit  non-pole
 behaviour. Using the available unpolarized
 data \cite{mersmann}, the authors of
 Ref.\cite{cwilkin07} found some non-direct indications
 of this specific phase behaviour of the s-wave amplitudes
 near the threshold. It is important to validate
 this interpretation by direct measurement of  energy dependence of the
 s-wave amplitudes, which requires polarization experiments.
 Some of them were discussed in Ref.\cite{cwilkin07} and
 it was assumed  that the
 tensor analyzing power $t_{20}$ and the spin correlation
 parameter $C_{y,y}$ even in collinear kinematics
 could give necessary information about the s-wave
 amplitudes.

 In this paper  polarization measurements are discussed 
 in detail. We consider a
 full set of spin observables including those
 which require measurement
 of polarizations of final particles.
 Moreover, in addition to the approximation by four spin amplitudes,
 used in Ref.\cite{cwilkin07}, we also investigate  the case with
 five amplitudes and the general case, which includes a 
 full set of six independent spin amplitudes of this reaction.
  We show that in collinear kinematics the complete polarization
 experiment is rather simple, but one cannot
 separate s- and  p- wave amplitudes in this regime.
 Therefore, to determine the desirable phase dependence
 of the s-wave amplitudes in respect of p-wave ones
 one has to perform experiments beyond collinear kinematics.
 For the cases of five and four independent amplitudes we consider
  complete polarization experiments, which allow one to
 determine all these amplitudes.
 We find that one has to measure spin-transfer coefficients in
 addition to analyzing powers and
 spin-correlation parameters to complete the experiment. 
 In the general case with six spin amplitudes the
 complete polarization experiment is too complicated
 to be really performed. Instead of performing it,
 we suggest to measure   
 few optimal spin observables which could provide
 new independent information on
 the  position of the quasi-bound pole. 

 In the next Section the spin structure of the transition matrix element
 is discussed. In section 3 all spin observables 
  are derived in the general form, assuming
 conservation of P-parity and angular momentum. In 
 Section 4  several 
 versions of complete polarization experiments are discussed.
 Section 5 is devoted to the minimal set of spin observables, measurements
 of  which should
 justify the position of the pole.
 A summary is given in the Conclusion. Some formulae for spin observables
 are given in the Appendix.

\section{The transition matrix element}

In terms of z-projections of spins of participating particles, 
 there exist twelve different spin  transitions
 in the reaction $pd\to~^3He\eta$.
 Due to P-parity and angular momentum conservation only six of
 them are independent. This can be seen, for example,
 in the following way.
 Let us denote the orbital momentum in the initial (final)
 state as $L_i$($L_f$) 
 and the total angular momentum as $J_f$ ($J_i$). 
 Then the  P-parity of the system
 is $\pi=(-1)^{L_i}=(-1)^{L_f+1}$.
 The spin of the  initial channel $S$,
 which is defined as a vector sum  of the
 deuteron and proton spins,
  takes two values $S=\frac{1}{2}$ and
 $S=\frac{3}{2}$. The sole relation $|L_i-L_f|=1$ is allowed
 for these values of $S$ due to angular momentum and parity
 conservation. 
  For the s-wave final state ($L_f=0$) only two transitions are allowed
  with $L_i=1$, $J_i=J_f=\frac{1}{2}$ and $S=\frac{1}{2}$ and  $\frac{3}{2}$.
 For the final p-wave state ($L_f=1$) there are five transitions, two for
  $L_i=0$ and three for $L_f=2$. For all higher partial waves
% (d- , f- ,$\dots$)
 ($L_f>1$) there are six independent
  transitions, three for $L_i=L_f+1$ and three for $L_i=L_f-1$.
 In a similar way one can
 find that for the given total angular momentum $J_i=J_f$
%\not =\frac{1}{2}$ 
 there exist 
%in a general case
 other six independent transitions.

 In the non-orthogonal basis used in Ref.\cite{cwilkin07}
 and in notations of that paper, the transition operator of
 the reaction $pd\to ~^3He\eta$ can be  written as
\begin{equation}
\label{f}
{\hat F}={\bfg \varepsilon}\cdot {\bf T}=
 A{\bfg \varepsilon}\cdot {\hat {\bf p}}_p+
iB[{\bfg \varepsilon}\times {\bfg \sigma}]\cdot {\hat {\bf p}}_p+
 C{\bfg \varepsilon}\cdot { {\bf p}}_\eta+
iD[{\bfg \varepsilon}\times {\bfg \sigma}]\cdot{ {\bf p}}_\eta 
+iE({\bfg \varepsilon}\cdot {\bf n})({\bfg \sigma}\cdot{\hat { \bf p}}_p)
+iF({\bfg \varepsilon}\cdot {\bf n})({\bfg \sigma}\cdot{ \bf p}_\eta),
\end{equation}
where $\bfg \sigma$ is the Pauli matrix, $\bfg \varepsilon$ is the
polarization vector of the deuteron, ${\hat {\bf p}_p}$ is the unit vector
along the proton beam direction, ${\bf p}_p$ and 
 ${\bf p}_\eta$ are the cms momenta of the proton and
the $\eta-$meson, respectively, and 
${\bf n}=[{\bf p}_\eta \times{\hat {\bf  p}}_p]$.
 We choose the coordinate system with the axes 
 OZ $\uparrow\uparrow {\bf p}_p$, OY $\uparrow\uparrow
%[{\bf p}_\eta \times{\hat {\bf p}}_p]$,
{\bf n} $,
 OX $\uparrow\uparrow [{\bf n}\times {\hat {\bf p}}_p]$.
% OX $\uparrow\uparrow [[{\bf p}_\eta \times{\bf p}_p]\times {\bf p}_p]$.
 Six independent terms in 
 Eq.(\ref{f}) correspond to six transitions discussed above and
 completely describe this reaction.
 As compared to Ref.\cite{cwilkin07}, 
 we also consider  two additional  terms, E and F,
 both of the  non-s-wave type.  
 The s-wave amplitudes are contained in the terms $A$ and $B$ only.
 These are the only terms which  do not
 disappear in the limit $p_\eta\to 0$ in Eq. (\ref{f}).
 In the near threshold region, the terms  $A$ and $B$ could contain mainly
 an admixture of p-waves,
 like $A'({\bf p}_p\cdot {\bf p}_\eta)(\bfg \varepsilon \cdot {\bf p}_p)$ and
 $iB'({\bf p}_p\cdot {\bf p}_\eta)[{\bfg \varepsilon}\times 
 {\bfg \sigma}]\cdot {\hat {\bf p}}_p$, respectively. These two p-wave
 terms and another p-wave amplitude  $E$ 
 correspond to the initial d-wave state,
 whereas the p-wave terms $C$ and $D$ are related to the initial s-wave state.
 This is in agreement with the discussion given in the beginning of
 this section.
 The last term  $F$  in Eq. (\ref{f}) corresponds to the 
 d-wave (and higher partial
 waves) in the  final state.

 The unpolarized cms cross section takes the form (see also below Eqs.
 (\ref{sigma06}) and (\ref{I6}))
 \begin{equation}
\label{sigma0}
d\sigma_0=\frac{p_\eta}{3\,p_p}I,
\end{equation}
here 
the factor $I$ is given as
\begin{eqnarray}
\label{I}
%I^{CW}=|C|^2q_x^2+2|D|^2q_x^2+|A+Cq_z|^2+2|B+Dq_z|^2,\\
I= |A|^2+2|B|^2+(|C|^2+2|D|^2)p_\eta^2+\nonumber \\
+2 Re(AC^*+2BD^*)p_\eta\cos{\theta_\eta}+ 
 (|F|^2p_\eta^2+|E|^2)p_\eta^2\sin^2{\theta_\eta}+\nonumber \\
+2 Re(DE^* -BF^*+ EF^*p_\eta\cos{\theta_\eta})p_\eta^2\sin^2{\theta_\eta}
\end{eqnarray}
with $\theta_\eta$ being the  angle between the vectors ${\bf p}_p$ and
${\bf p}_\eta$.
 If one puts $F=0$ and $E=0$, 
 Eq. (\ref{I}) coincides with Eq.(4)
 from Ref.\cite{cwilkin07} and gives linear dependence 
 in $\cos{\theta_\eta}$ for the differential
 cross section, observed  in the existing
 data at $Q<11$ MeV \cite{mersmann,smyrski}.
 In view of this observation, one may conclude the following.
 Firstly, 
 the contribution of two last terms in Eq.(\ref{I}),
 which are proportional to $\sin^2{\theta_\eta}$, can be considered
 as negligible
 in the unpolarized cross section at $Q$ less than $\approx 10$ MeV.
 The latter  gives grounds for using only
 four independent amplitudes and neglecting $E$ and $F$.
 This approximation, adopted in Ref.\cite{cwilkin07}, will also be 
 studied
 in this paper  from the point of view of performing a
 complete polarization experiment.
 Secondly, the amplitudes $A$ and $B$
  do not contain a sizeable contribution of p-waves 
 at  $Q<11$ MeV \cite{cwilkin07} \footnote{I am thankful to C. Wilkin
 for explanation of this
 feature.}.
 Thus, near the threshold the amplitudes $A$ and $B$ are of the
 s-wave type and,
 therefore, are expected to contain the quasi-bound pole. On the other hand,
 all other amplitudes ($C$, $D$, $E$, and $F$) are expected to have smooth
 $p_\eta$ dependence near the pole.
  These assumptions will be essential 
 in section \ref{OO}, where the position of the quasi-bound pole on the
 momentum plane is considered.  One should note that the d-wave
 term $F$ is
 most likely  negligible as compared to the s- and p-wave terms
 in the near threshold region due to the centrifugal
 barrier. However, 
 the  term $E$  could be non-negligible in spin observables.
 Therefore, below we will also
 consider the case with five spin 
 amplitudes ($A$, $B$, $C$, $D$, $E$)
 and the general case with the full set of six spin amplitudes.

\section{Observables}

Our strategy is the following. First, we derive a full set of spin observables
for the general case of six independent spin amplitudes. After that
 the formalism for a particular
 case with  five or four amplitudes can be obtained straightforwardly
 from the general formalism. Furthermore, having the full set of spin
 observables,
 one can
 find a solution  for  complete polarization experiments.

\subsection{ General case}

 Using Eq. (\ref{f}), one can express  any spin observables in terms of
 invariant spin amplitudes. With this aim, we present  
 the Cartesian components of the vector-operator ${\bf T}$, defined by 
 Eq. (\ref{f}), as  follows
 \begin{eqnarray}
 \label{T4-pwave}
T_x=M_1+M_2\sigma_y, \, 
T_y=M_3\sigma_x+M_4\sigma_z, \,
T_z=M_5+M_6\sigma_y,
\end{eqnarray}
where 
\begin{eqnarray}
\label{mi-4pwave}
M_1=Cq_x,\, M_2=i(B+Dq_z),\, M_3=-i(B+Dq_z-Fq_x^2),\, \nonumber \\
M_4=i(D+E+Fq_z)q_x,\, M_5=A+Cq_z,\, M_6=-iDq_x.
\end{eqnarray}
Here $q_x$ and $q_z$ are determined  as
\begin{eqnarray}
  \label{qxqz}
  q_x=-p_\eta sin\theta_\eta,\,
   q_z=p_\eta cos\theta_\eta.
  \end{eqnarray}

The unpolarized cross section can be written as
\begin{eqnarray}
\label{sigma06}
d\sigma_0=\frac{1}{6}\frac{p_\eta}{p_p}\sum_{\alpha}TrT_\alpha T^+_\alpha
=\frac{1}{3}\frac{p_\eta}{p_p}I,
\end{eqnarray}
where $\alpha=x,y,z$ and $I$ has the  following form:
\begin{equation}
\label{I6}
I=\frac{1}{2}\sum_{\alpha}TrT_\alpha T^+_\alpha=\sum_{i=1}^6|M_i|^2.
\end{equation}

We use below definitions and notations of
 Ref. \cite{ohlsen} for spin observables.
The tensor analyzing powers  of the deuteron take the form
\begin{eqnarray}
 \label{ayy6}
A_{yy}=1-\frac{3}{I}\left \{|M_3|^2+|M_4|^2\right \}, \\
 \label{axx6}
A_{xx}=1-\frac{3}{I}\left \{|M_1|^2+|M_2|^2\right \}, \\
 \label{azz6}
A_{zz}=1-\frac{3}{I}\left \{|M_5|^2+|M_6|^2\right \},\\ 
\label{axz6}
-\frac{I}{3}A_{xz}=Re(M_1M_5^*+M_2M_6^*).
\end{eqnarray}
The vector analyzing powers are
\begin{eqnarray}
\label{ayd6}
\frac{I}{2}A_y^d=Im\left \{M_5M_1^*+M_6M_2^*\right \},\\
\label{ayp6}
\frac{I}{2}A_y^p=Re\left \{M_1M_2^*+M_5M_6^*\right \}-ImM_3M_4^*.
\end{eqnarray}
The spin transfer coefficients $K_i^{j'}(p)$, describing polarization
 transfer from the initial proton  to the $^3He$ nucleus, can be
 written as 
\begin{eqnarray}
\label{kyyp6}
IK_y^{y'}(p)=|M_1|^2+|M_2|^2-|M_3|^2-|M_4|^2+ |M_5|^2+|M_6|^2,\\
\label{kxxp6}
IK_x^{x'}(p)=|M_1|^2-|M_2|^2+|M_3|^2-|M_4|^2+ |M_5|^2-|M_6|^2,\\
\label{kzzp6}
IK_z^{z'}(p)=|M_1|^2-|M_2|^2-|M_3|^2+|M_4|^2+ |M_5|^2-|M_6|^2,\\
\label{kxzp6}
-\frac{I}{2}K_x^{z'}(p)=Im\left \{M_1M_2^*+M_5M_6^*\right \}-ReM_3M_4^*,\\
\label{kzxp6}
-\frac{I}{2}K_z^{x'}(p)=Im\left \{M_2M_1^*+M_6M_5^*\right \}-ReM_3M_4^*.
 \end{eqnarray}

The spin transfer coefficients $K_j^{i'}(d)$, describing  polarization transfer
 from the initial deuteron to the $^3He$ nucleus,  are 
\begin{eqnarray}
\label{kyyd6}
\frac{I}{2}K_y^{y'}(d)=Im(M_6M_1^*+M_5M_2^*),\nonumber\\
\label{kxxd6}
\frac{I}{2}K_x^{x'}(d)=ImM_3M_5^*-ReM_4M_6^*,\nonumber\\
\label{kzzd6}
\frac{I}{2}K_z^{z'}(d)=ImM_1M_4^*-ReM_2M_3^*,\nonumber\\
\label{kzxd6}
\frac{I}{2}K_z^{x'}(d)=ImM_1M_3^*+ReM_2M_4^*,\nonumber\\
\label{kxzd6}
\frac{I}{2}K_x^{z'}(d)=ImM_4M_5^*+ReM_3M_6^*.
\end{eqnarray}

 The proton-deuteron spin correlation parameters are
% for the vector polarized deuteron are
%we find
\begin{eqnarray}
\label{cyy6}
-\frac{I}{2}C_{y,y}=ImM_2M_5^*+ImM_1M_6^*,\nonumber\\
\label{czz6}
\frac{I}{2}C_{z,z}=ImM_1M_4^*+ReM_2M_3^*,\nonumber\\
\label{cxx6}
\frac{I}{2}C_{x,x}=ImM_3M_5^*+ReM_4M_6^*,\nonumber\\
\label{cxz6}
\frac{I}{2}C_{x,z}=ImM_1M_3^*-ReM_2M_4^*,\nonumber\\
\label{czx6}
\frac{I}{2}C_{z,x}=ImM_4M_5^*-ReM_3M_6^* 
\end{eqnarray}
 for the vector polarized deuteron and
% and for the tensor polarized deuteron 
%The proton-deuteron spin-tensor correlation parameters can be presented as 
\begin{eqnarray}
\label{cyyy6}
C_{y,yy}=A_y^p+\frac{6}{I}ImM_3M_4^*,\nonumber\\
\label{cyyy6m}
\frac{I}{2}C_{y,yy}=Re(M_1M_2^*+M_5M_6^*)+2ImM_3M_4^*,\nonumber\\
\label{cyxx6}
C_{y,xx}=A_y^p-\frac{6}{I}ReM_2M_1^*,\nonumber\\
\label{cyxx6m}
\frac{I}{2}C_{y,xx}=Re(M_5M_6^*-2M_2M_1^*)-ImM_3M_4^*,\nonumber\\
\label{cyzz6}
C_{y,zz}=A_y^p-\frac{6}{I}ReM_5M_6^*,\nonumber\\
\label{cyzz6m}
\frac{I}{2}C_{y,zz}=Re(M_1M_2^*-2M_5M_6^*)-ImM_3M_4^*,\nonumber\\
\label{cyxz6}
-\frac{I}{3}C_{y,xz}=Re(M_2M_5^*+M_1M_6^*),\nonumber\\
\label{cxyz6}
-\frac{I}{3}C_{x,yz}=ReM_3M_5^*-ImM_4M_6^*,\nonumber\\
\label{czyz6}
-\frac{I}{3}C_{z,yz}=ReM_4M_5^*-ImM_6M_3^*,\nonumber\\
\label{cxxy6}
-\frac{I}{3}C_{x,xy}=ReM_1M_3^*-ImM_4M_2^*,\nonumber\\
\label{czxy6}
-\frac{I}{3}C_{z,xy}=ReM_1M_4^*-ImM_2M_3^*,\nonumber\\
C_{x,xx}=C_{z,xx}=C_{z,zz}=C_{x,zz}
%=\nonumber\\
=C_{y,xy}=C_{y,yz}=C_{x,xz}=C_{z,xz}=0
\end{eqnarray}
for the tensor polarized deuteron.
Polarization transfer  from the tensor polarized
 deuteron to the $^3He$ nucleus is described by the following 
 observables:
\begin{eqnarray}
\label{kyyypy}
 K_{yy}^{y'}=P_y^{h}+\frac{6}{I}ImM_4M_3^*,\nonumber\\
\label{kyyy}
\frac{I}{2}K_{yy}^{y'}=Re(M_1M_2^*+M_5M_6^*)-2ImM_4^*M_3,\nonumber\\
\label{kxxypy} 
K_{xx}^{y'}=P_y^{h}-\frac{6}{I}ReM_2M_1^*,\nonumber\\
\label{kxxy}
\frac{I}{2}K_{xx}^{y'}=Re(M_5M_6^*-2M_1M_2^*)+ImM_3M_4^*,\nonumber\\
\label{kzzypy}
K_{zz}^{y'}=P_y^{h}-\frac{6}{I}ReM_5M_6^*,\nonumber\\
\label{kzzy}
\frac{I}{2}K_{zz}^{y'}=Re(M_1M_2^*-2M_5M_6^*)+ImM_3M_4^*,\nonumber\\
\label{kyzx}
-\frac{I}{3}K_{yz}^{x'}=ReM_3M_5^*+ImM_4M_6^*,\nonumber\\
\label{kxyz}
-\frac{I}{3}K_{xy}^{z'}=ReM_1M_4^*+ImM_2M_3^*,\nonumber\\
\label{kxzy}
-\frac{I}{3}K_{xz}^{y'}=Re(M_2M_5^*+M_1M_6^*),\nonumber\\
\label{kyzz}
-\frac{I}{3}K_{yz}^{z'}=ReM_4M_5^*+ImM_6M_3^*,\nonumber\nonumber\\
\label{kxyx}
-\frac{I}{3}K_{xy}^{x'}=ReM_1M_3^*+ImM_2^*M_4,\nonumber\nonumber\\
\label{kijknull}
K_{xx}^{x'}=K_{xx}^{z'}=K_{zz}^{z'}=K_{zz}^{x'}=
K_{xy}^{y'}=K_{yz}^{y'}=K_{xz}^{x'}=K_{xz}^{z'}=0,
\end{eqnarray}
where $P_y^{\tau}$ is the polarization of the final $~^3He$ nucleus
 for the unpolarized beam
 and target, which can be written as
\begin{eqnarray}
\label{pytau}
\frac{I}{2}P_y^{\tau}=Re(M_1M_2^*+M_5M_6^*)+ImM_3M_4^*,\,
 P_x^{\tau}=P_z^{\tau}=0.
\end{eqnarray}

One can also find the following relations:
\begin{eqnarray}
\label{kayy6}
3K_y^{y'}(p)=2A_{yy}+1,\, \\
\label{kcyy6}
K_y^{y'}(d)=C_{y,y}.
\end{eqnarray}

\subsection{ Five independent  amplitudes}
 Assuming $F=0$ in Eq. (\ref{f}), one has $M_2=-M_3$.
 In this case the independent amplitudes
 in Eqs.(\ref{sigma06})-(\ref{pytau})
 are
\begin{eqnarray}
\label{m6tom5}
M_1=Cq_x,\,M_2=i(B+Dq_z),\,M_3=-i(B+Dq_z), \,\nonumber \\
M_4=i(D+E)q_x, \,M_5=A+Cq_z,\, M_6=-iDq_x.
\end{eqnarray}

\subsection{ Four independent  amplitudes}

If one puts $E=F=0$ in Eq.(\ref{f}), then one has 
 $M_2=-M_3$ and $M_4=-M_6$.
In this case the spin structure of Eq. (\ref{f})  
coincides with that used in Ref. \cite{cwilkin07}.
 Thus, the following replacements in
 Eqs.(\ref{sigma06})-(\ref{pytau})
%(\ref{realtions26}) 
\begin{eqnarray}
\label{m6tom4}
M_1=Cq_x,\,M_2=i(B+Dq_z),\,M_3=-M_2, \,M_4=iDq_x,\,M_5=A+Cq_z,\, M_6=-M_4
\end{eqnarray} 
reduce the general case to the approximation by four spin amplitudes. 
The spin observables in terms of the amplitudes $A$, $B$, $C$ and $D$ 
for this case are given in the Appendix.

\section{Complete polarization experiment}

\subsection{Collinear kinematics}

In collinear kinematics ($q_x=0$) one has $M_1=M_4=M_6=0$ and $M_3=-M_2$,
as  follows from Eq. (\ref{mi-4pwave}).
Therefore, only two  spin
 amplitudes, $M_2=M_2^{coll}$ and $M_5=M_5^{coll}$, completely describe 
 the reaction.  
 These amplitudes
 can be determined from the measurement of the following  four observables:
\begin{eqnarray}
\label{sigcollin6}
d\sigma_0^{coll}=\frac{1}{3}\frac{p_\eta}{p_p}\left 
(2|M_2^{coll}|^2+|M_5^{coll}|^2\right )=\frac{1}{3}\frac{p_\eta}{p_p}I^{coll},\\
\label{azzcoll6}
A_{zz}^{coll}=1-\frac{3}{I^{coll}}|M_5^{coll}|^2,\\
\label{cyycoll6}
-\frac{I^{coll}}{2}C_{y,y}^{coll}=Im M_2^{coll} M_5^{coll *},\\
\label{cyxzcoll6}
-\frac{I^{coll}}{2}C_{y,xz}^{coll}=Re M_2^{coll} M_5^{coll *}.
\end{eqnarray}
 From Eqs. (\ref{sigcollin6}) and (\ref{azzcoll6}) one can find  moduli
of these amplitudes, whereas  the relative phase $\phi_{25}$ can be found from
$\sin{\phi}_{25}$ and $\cos{\phi}_{25}$   given
 by Eqs. (\ref{cyycoll6}) and (\ref{cyxzcoll6}), respectively.
Here the relative phase $\phi_{ij}$ of the amplitudes $M_i$ and
$M_j$ is defined as $ Re M_iM_j^*=|M_i||M_j|\cos{\phi_{ij}}$,
 $Im M_iM_j^*=|M_i||M_j|\sin{\phi_{ij}}$.
 In the general case,   $M_2^{coll}$
 and $M_5^{coll}$ are the following  combinations of the
 s- and p-wave amplitudes:
 $M_5^{coll}=A\pm Cp_\eta$
 and $M_2^{coll}=B\pm Dp_\eta$, where the $\pm$sign refers  to
 forward and backward production of the $\eta$ meson, respectively.
 Obviously, 
in collinear kinematics one cannot separate s- and p-wave amplitudes.
 For example, one can measure $t_{20}$ for forward and
 backward production, as  suggested in Ref. \cite{cwilkin07},
 which gives 
 $Re A^*C/(|A|^2+p_\eta^2|C|^2)$  and $Re B^*D/(|B|^2+p_\eta^2|D|^2)$
(see Eq.(11) in Ref.\cite{cwilkin07}).
 But  the desirable  values of $cos\phi_{AC}$
 and $cos\phi_{BD}$ cannot be extracted from
 these measurements 
\footnote{At the limiting point
 $p_\eta=0$, where
 the p-wave amplitudes  vanish, 
 the separation of  two s-wave
 amplitudes can be performed
 exactly.}.  
 Measurement of $C_{y,y}$  in collinear kinematics
 gives $Re AB^*$ with an admixture of
 $Re(A^*D+BC^*)p_\eta+Re C^*Dp_\eta^2$ and, therefore,
  cannot help to disentangle the moduli and phases of
 the amplitudes. 
 In order to separate the amplitudes $A$ and  $B$ (or  $C$ and  $D$), 
 measurements beyond  collinear kinematics  are required.

\subsection{ Four spin amplitudes}
\label{CPE4}
\subsubsection{ Involving only  analyzing powers and spin correlations}
\label{CPE4as}
%In this subsection we suppose that spin transfer coefficients $K_i^j$
% are not measured.
 When neglecting the terms $E$ and $F$  in Eq. (\ref{f}),
 one can find from Eqs. (\ref{cyy}), (\ref{cxx}) and (\ref{czz}) the only
 relation which contains moduli of the amplitudes and does not involve 
their phases:
\begin{eqnarray}
\label{rel1}
\frac{I}{2}\left [C_{y,y}-C_{x,x}+C_{z,z}\right ]=\frac{I}{3}(1-A_{yy})=
|D|^2q_x^2+|B+Dq_z|^2.
\end{eqnarray} 
However,  this relation 
 does not provide new information as compared to $A_{yy}$.
Furthermore, taking into account the relation
\begin{equation}
\label{rel2}
 A_{xx}+A_{yy}+A_{zz}=0,
\end{equation} 
 one can find from 
 Eqs.(\ref{sigma06}), (\ref{I6}), (\ref{I4})-(\ref{azz}),
 (\ref{cyy})-(\ref{czz}), and  (\ref{rel1}) that 
  the unpolarized  cross section $d\sigma_0$,
 tensor analyzing powers $A_{ij}$ and  spin correlation
 coefficients $C_{i,j}$ and $C_{i,jk}$ allow one to derive
 only three independent linear
 equations containing  the squared  moduli of the following  four
 amplitudes:
 $M_a=A+Cq_z,\, M_{b}=B+Dq_z,\, C$ and $D$. Therefore, one cannot determine
 the moduli independently of the phases of these amplitudes 
% Instead, one has to determine the moduli and phases simultaneously.
and  has to find all of them  simultaneously. For this aim,
 one can use  
 nine observables $d\sigma_0$, $A_{xx}$, $A_{yy}$, $A_{xz}$,
 $C_{y,y}$, $C_{x,x}$,
 $C_{z,z}$, $C_{x,z}$, and  $C_{z,x}$, which  give nine (nonlinear) 
equations (\ref{I4}), (\ref{axx}), (\ref{ayy}), (\ref{axz}),
% (\ref{cyy}),(\ref{cxx}), (\ref{czz}), (\ref{cxz}) and (\ref{czx}),
 (\ref{cyy}) -- (\ref{czx}),
 connecting 
 four moduli  and six  cosines of  phases  $\cos\phi_{CD}$,
$\cos\phi_{ab}$, $\cos\phi_{Ca}$, $\cos\phi_{Cb}$, $\cos\phi_{Da}$,
$\cos\phi_{Db}$. Due to Eq. (\ref{rel1}), only eight equations are
 independent. 
Furthermore, since only three relative phases are 
independent, one
 should  add to these system the following  three linear
 equations connecting six relative  phases: $\phi_{Cb}-\phi_{Ca}=\phi_{ab}$,
$\phi_{Db}-\phi_{Da}=\phi_{ab}$, $\phi_{Cb}-\phi_{Db}=\phi_{CD}$.
  However, in order to really use these relations
 one also has to employ  equations containing sines of the phases.
Therefore, one should take, for example,
six observables  $A_y^d$, $A_y^p$, $C_{y,yy}$, $C_{y,zz}$,
 $C_{x,yz}$, and $C_{z,xy}$   given by Eqs. (\ref{ayd}), (\ref{ayp})
 (\ref{cyyy2}), (\ref{cyzz2}),
  (\ref{cxyz}) and  (\ref{czxy}), respectively. 
  In total,  one has  17 equations, which contain 14  spin observables
 and 16 unknown variables (four moduli, six cosines and six sines of
 the phases). Most likely, a
 solution to this system of equations can be found numerically, but
 this version of complete polarization experiment is too
 cumbersome.

\subsubsection{Involving spin-transfer coefficients}
\label{CPE4st}
 The problem  becomes much simpler 
 if one can measure spin transfer coefficients.
  A possible solution is the following. The moduli
 of the amplitudes $M_a=A+Cq_z$, $M_b=B+Dq_z$, $D$ and $C$
 can be found from
 the measurement of
% the following four observables
 $d\sigma_0$, $A_{yy}$, $A_{xx}$
 and $K_x^{x'}(p)$ as
\begin{eqnarray}
\label{c2}
|C|^2q_x^2=\frac{I}{3}\left [ 2A_{yy}-A_{xx}-1 \right ]+
\frac{I}{2}(1-K_x^{x'}(p)),\\
\label{m2}
|B+Dq_z|^2=\frac{2I}{3}(1-A_{yy})-\frac{I}{2}(1-K_x^{x'}(p)),\\
\label{d2}
|D|^2q_x^2=\frac{I}{2}(1-K_x^{x'}(p))-\frac{I}{3}(1-A_{yy}),\\
\label{m1}
|A+Cq_z|^2=\frac{I}{3}(2+A_{xx})-\frac{I}{2}(1-K_x^{x'}(p)).
\end{eqnarray}

In order to completely determine four complex
amplitudes three relative phases
 have to be measured (note  that the
common phase is non-measurable
and one can put it  equal to zero).
  A possible  solution for the relative phases of the
 amplitudes $C$, $M_b$, $D$, and $M_a$, whose
% $M_1$, $M_2$, $M_4$, and $M_5$, whose
  moduli are 
 given in Eqs.(\ref{c2}),  (\ref{m2}), (\ref{d2}) and (\ref{m1})  
 is to measure
 $K_x^{z'}(p)$, $K_z^{x'}(p)$ and $K_x^{z'}(d)$. 
 It gives  $\cos{\phi_{Db}}$, $\cos{\phi_{Cb}}$  and $\cos{\phi_{Da}}$, 
 as it follows
 from the following relations:
\begin{eqnarray}
\label{kplusk}
 I[K_x^{z'}(p)+K_z^{x'}(p)]=-2Re D^*(B+Dq_z)q_x, \nonumber \\
\label{kminuskd}
 I[K_x^{z'}(p)-\frac{1}{2}K_x^{z'}(d)]=q_x Re (B+Dq_z)C^*,\nonumber\\
\label{kminusk}
 I[K_x^{z'}(p)-K_z^{x'}(p)]=
-2q_x\left \{Re (B+Dq_z)C^*-(A+Cq_z)D^*\right \}.
\end{eqnarray}
 In order to find
% $\sin{\phi_{D1}}$,  $\sin{\phi_{D2}}$ and  $\sin{\phi_{C2}}$,
 $\sin{\phi_{Da}}$,  $\sin{\phi_{Db}}$ and  $\sin{\phi_{Cb}}$,
 one has to measure
% $A_y^p$,  $A_y^d$ and $C_{y,zz}$,
  $C_{y,yy}$, $C_{y,xx}$ and  $A_y^p$,
 as  follows from Eqs.
% (\ref{ayd}), (\ref{cyzz}) and (\ref{ayp}).
 (\ref{cyyy}), (\ref{cyzz}),  (\ref{cyxx})
 and the following
 relation: $C_{y,yy}+C_{y,xx}+C_{y,zz}=0$.
 Thus, in
 this version one has to perform ten accurate
  measurements. 

The first equation in Eq.(\ref{rel1}) can be considered as a necessary 
criterion for applicability of the approximation with
four spin amplitudes in the matrix element. 
 
\subsection{Five spin amplitudes}
\label{CPE5}

 Neglecting the d-wave amplitude $F$ in Eq.(\ref{f}), one has  
 the relation $M_2=-M_3$. For this case moduli of five 
amplitudes can be found by measurement of 
$d\sigma_0$, $A_{xx}$, $A_{yy}$, $K_x^{x'}(p)$ and $K_z^{z'}(p)$:
\begin{eqnarray}
\label{moduli5-1}
|M_1|^2=\frac{I}{4}\left \{K_z^{z'}(p)-K_x^{x'}(p)\right \}+
\frac{I}{6}\left ( 1+A_{yy}-2A_{xx}\right ),\\
\label{moduli5-2}
|M_2|^2=-\frac{I}{4}\left \{K_z^{z'}(p)-K_x^{x'}(p)\right \}+
\frac{I}{6}\left ( 1-A_{yy}\right ),\\
\label{moduli5-4}
|M_4|^2=\frac{I}{4}\left \{K_z^{z'}(p)-K_x^{x'}(p)\right \}+
\frac{I}{6}\left ( 1-A_{yy}\right ),\\
\label{moduli5-5}
|M_5|^2=\frac{I}{2}\left \{ \frac{1}{3} +\frac{2}{3}A_{xx}+ K_x^{x'}(p)
\right \},\\
\label{moduli5-6}
|M_6|^2=\frac{I}{2}\left \{ \frac{1}{3} +\frac{2}{3}A_{yy}- K_x^{x'}(p)
\right \}.
\end{eqnarray}
 Among the available five phases of these amplitudes only four are independent.
In order to determine two independent phases, one can measure, for example,
the following four observables:
  $K_x^{z'}(d)$, $C_{z,x}$,  $C_{z,yz}$ and 
 $K_{yz}^{z'}$. Indeed, one can see from Eqs. (\ref{kxzd6}), (\ref{czx6}),
  (\ref{czyz6}) and (\ref{kyzz}) that the above observables
 completely determine the phases $\phi_{45}$ and $\phi_{63}$.
The other two phases $\phi_{13}$ and $\phi_{42}$ can be determined
by measurement of  $C_{x,z}$, $K_z^{x'}(d)$, $C_{x,xy}$ and
 $K_{xy}^{x'}$. Thus, one needs 13 accurate measurements in this variant.

\subsubsection{Reduction to four spin amplitudes}
\label{CPE4t}
  The solution obtained
%CHECK potapov
 for the case of  five spin  amplitudes
 can be reduced to  four 
 amplitudes. This gives a new  solution in addition to that
 found  in section \ref{CPE4st}. 
 Indeed, when neglecting the p-wave amplitude $E$ and d-wave amplitude
 $F$, one has $M_2=-M_3$ and $M_4=-M_6$. In this case one finds 
from Eqs. (\ref{moduli5-4})
 and (\ref{moduli5-6}) the following relation:
\begin{equation}
\label{kxxkzzayy}
 K_x^{x'}(p)+K_z^{z'}(p)=2A_{yy}.
\end{equation}
 This relation can be used as a necessary condition for validity of the
 approximation by four spin amplitudes. Using Eq. (\ref{kxxkzzayy})
 one can find from  Eqs. (\ref{moduli5-1})-(\ref{moduli5-6}) that four
 observables  $d\sigma_0$, $A_{xx}$, $A_{yy}$ and $K_x^{x'}(p)$  
 determine four moduli $|M_1|$, $|M_2|$, $|M_4|$ and $|M_5|$. Taking into
 account Eqs.(\ref {m6tom4}), one can see that these formulae  
 coincide with that given by Eqs.(\ref{c2})-(\ref{m1}). 
 Three
 phases $\phi_{14}$, $\phi_{34}$ and $\phi_{56}$ can be determined by the
% following
 observables $C_{z,z}$, $C_{y,yy}$, $C_{z,xy}$, $A_y^p$,
 $K_x^{z'}(p)$ , and $K_z^{x'}(p)$ as it follows from  Eqs. (\ref{czz6}),
(\ref{cyyy6}), 
%CHECK
%(\ref{czxy6}), 
(\ref{ayp6}), (\ref{kxzp6}) and (\ref{kzxp6}).
 Finally, using the relation 
$\phi_{56}=\phi_{45}+\pi$, one can see that four amplitudes
 ($M_1$, $M_2$, $M_4$, $M_5$) 
 are completely 
 determined  by ten observables, 
 $d\sigma_0$, $A_{xx}$, $A_{yy}$, $K_x^{x'}(p)$, 
 $C_{z,z}$, $C_{z,xy}$, $C_{y,yy}$, $A_y^p$, $K_z^{x'}(p)$ and $K_x^{z'}(p)$.
 Three of these observables 
 (e.g., $C_{z,xy}$, $K_z^{x'}(p)$ and $K_x^{z'}(p)$)
 can be measured roughly.

\subsection{Full set of spin amplitudes}
\label{CPE6}
 In the general case, when all six spin amplitudes $M_i$ ($i=1,\dots, 6$)
 are included in the  consideration, 
 there is no simple way to determine moduli of
 these amplitudes. Indeed, there are seven linear equations for six squared
 moduli of these amplitudes given by  Eqs. (\ref{I6}), 
(\ref{ayy6}), (\ref{axx6}),
 (\ref{azz6}), (\ref{kyyp6}), (\ref{kxxp6}), (\ref{kzzp6}),
 which do not involve any  phases.
 However, only five of them are linearly independent due to
 Eqs.(\ref{kayy6}) and (\ref{rel2}). Therefore, in the general case
 in contrast to
 the case with four and five  spin amplitudes,
 considered in sections \ref{CPE4st} and \ref{CPE5}, respectively,
 one cannot determine the moduli of the amplitudes $|M_i|$ ($i=1, \dots, 6$)
 independently of their
 phases. It means  that the complete polarization experiment is too
 complicated in this case.  Theoretically it suggests simultaneous
 determination of the
 moduli and phases of the all amplitudes. 
 Therefore, one has to deal with a large 
 number of (nonlinear) equations, as  in  
 subsection \ref{CPE4as}.  Furthermore, here one has six amplitudes
 instead of four and, therefore, the problem becomes even more complicated.
 Thus, we will no longer consider  the 
 complete polarization experiment for this general case.

\section{Specific polarization observables}
\label{OO}
The complete polarization experiment provides full information about the
reaction, including possible  presence and position of the quasi-bound pole.
However, in order to find solely the 
specific phase behaviour of the s-wave
 amplitudes caused by
 the $~^3He-\eta$ pole, one should select only
 those  observables which are  most informative in this respect.
 In this connection one should note that at $q_z=0$
 the amplitudes 
 $M_2$, $M_3$ and $M_5$ are of the s-wave type and, therefore, are
 expected to contain 
 the $^3He-\eta$ pole. On the other hand,
 the p- and d- wave amplitudes $M_1$, $M_4$ and $M_6$ are of the 
 non-pole type with
  smooth $p_\eta$ dependence. We show below that
%CHECK referee-II in some cases
%% it is possible to measure
 one can get 
 a product of s-wave  and p- wave amplitudes
 without performing a complete polarization experiment.
 Indeed, as  follows from Eqs.  (\ref{kzxd6}), (\ref{cxz6}),
 (\ref{cxxy6}) and (\ref{kxyx}),
 the observables  $K_z^{x'}(d)$, $C_{x,z}$, $C_{x,xy}$ and
 $K_{xy}^{x'}$ determine  the following complex numbers 
 $M_1^*M_3$ and $M_4^*M_2$: $ReM_2M_4^*=\frac{I}{4}[K_z^{x'}(d)-C_{x,z}]$,
 $Im M_2M_4^*=\frac{I}{6}[K_{xy}^{x'}-C_{x,xy}]$,
 which  can be written as
\begin{eqnarray}
\label{z1}
Z_1=M_2M_4^*=q_x\left \{B(D^*+E^*)+[D(E^*+D^*)+(B+D)F^*)]q_z\right \},\\
\label{z2}
Z_2=M_3M_1^*=-iq_x\left \{BC^*+DC^*q_z-FC^*q_x^2\right \}.
\end{eqnarray}
 At $q_z=0$, both of them are products of one s-wave and
 one p-wave amplitude \footnote {The contribution  of the d-wave term $Fq_x^2$
 to Eq. (\ref{z2}) 
%$M_3$
 is   negligible in the near threshold region.} and,
 therefore, are 
 expected to have the  pole form of Eq. (\ref{pole}).

% Similarly,
 Likewise, one can find from Eqs.(\ref{kxzd6}), (\ref{czx6}),
  (\ref{czyz6}) and (\ref{kyzz})
  that four observables $K_x^{z'}(d)$, $C_{z,x}$,  $C_{z,yz}$ and 
 $K_{yz}^{z'}$ completely
 determine the following complex numbers:
\begin{eqnarray}
\label{z3z4}
Z_3=M_5M_4^*=-i q_x\left \{A(D^*+E^*)+(AF^*+CD^*+CE^*)q_z+CF^*q_z^2\right \},\\
Z_4=M_3M_6^*=\left \{BD^*+|D|^2q_z-FD^*q_x^2\right \}q_x,
\end{eqnarray}
 which   are also expected to have the same pole form at $q_z\to 0$.
 Thus, even without knowing the moduli of the amplitudes
$|M_i|$ ($i=1,\dots, 6$), one can single out
 the pole dependence  (\ref{pole})  of the measured complex numbers $Z_k$
 ($k=1,\dots, 4$).

Furthermore, using Eq. (\ref{pole}), where $f(p_{\eta})=Z_k(p_{\eta})$, and
 taking real and imaginary parts of $(p_{\eta}-ip_0)Z_k$,
 one can find
% from Eq.(\ref{pole})
 near the pole
\begin{eqnarray}
\label{rezi}
(p_\eta+Imp_0)ReZ_k+Rep_0Im{Z_k}=Re{\xi} \approx const,\\
\label{imzi}
Rep_0Re{Z_k}-(p_\eta+Imp_0)ImZ_k=-Im{\xi} \approx const;
\end{eqnarray}
here $k=1,\dots, 4$ and {\it const} means independence of $p_\eta$.  
The magnitudes
% absolute values
 of $Im p_0$ and $Rep_0$  are restricted  by 
%the fit to
the quasi-bound state energy $Q_0=-p_0^2/2\mu$, determined from   
the existing unpolarized data
 \cite{mersmann}:
\begin{equation}
\label{absp0}
%(Re p_0)^2+(Imp_0)^2=|p_0|^2.
2\mu ReQ_0=(Imp_0)^2- (Rep_0)^2, \, \mu |ImQ_0|=|Re p_0| |Imp_0|,
\end{equation} 
 where $\mu$ is the reduced mass of the $^3He-\eta$ system.
  Eqs. (\ref{rezi}), (\ref{imzi})
% with the restriction by 
 and Eq. (\ref{absp0}) can be used
%CHECK referee-II
% justify
 for more precise determination of the magnitudes of
 $Imp_0$ and $Re p_0$ and
 can give  their signs. Knowledge of the sign of the real part
 of $p_0$ allows one to conclude
 whether the $^3He-\eta$ state is quasi-bound ($Rep_0>0$) or anti-bound 
 ($Rep_0<0$) \footnote{At the pole $p_\eta=ip_0$ defined by Eq. (\ref{pole}),
 the radial part  of the s-wave function of the $^3He-\eta$ relative motion
 has the asymptotic form of $\sim \exp{(ip_\eta r)}/r=\exp{[(-Rep_0-iIm p_0)r]}
/r$. This function vanishes  at $r\to \infty$ (bound) for $Rep_0>0$
 and increases infinitely (anti-bound) for $Rep_0<0$.}.
%Note that our definition of $p_0$ in Eq. (\ref{pole})
% differs from
% \cite{cwilkin07} by the imaginary unit $i$.}.  

 Measurement of spin transfer coefficients is a complicated experimental
 problem. In this connection one should note that measurement of only
 two observables $A_y^d$ and $A_{xz}$
 completely determines the following  complex number
\begin{eqnarray}
\label{z5}
Z_5=M_1M_5^*+M_2M_6^*= 
\left \{CA^*-BD^*+(|C|^2-|D|^2)q_z\right\}q_x.
\end{eqnarray}
 This can be seen from the relations
 $ReZ_5=-\frac{I}{3}A_{xz}$ and
 $ImZ_5=-\frac{I}{2}A_y^d$, following  from
 Eqs.(\ref{axz6}) and (\ref{ayd6}), respectively.
 This result was found in Ref. \cite{cwilkin07}, but within  the approximation
 by four spin amplitudes. We should stress that Eq. (\ref{z5}) is also
 valid in the general case of six independent amplitudes.
 In the pole region, $Z_5$ is a linear combination of the 
 $(p_\eta-ip_0)^{-1}$ and $(p_\eta+ip_0^*)^{-1}$ terms and, therefore,
 Eqs. (\ref{rezi}) and (\ref{imzi}) are not valid for $Z_5$. Nevertheless,
 this observable
 can be used as a new  crucial test for the existing models of the
 $pd\to ~^3He\eta$
 and $~^3He\eta\to~^3He\eta$ reactions.   
As follows from the $pd\to~^3He\eta$ data 
 on $t_{20}$ \cite{berger},  $|A|\approx |B|$.
One can see  from Eq. (\ref{z5}), that studying the $q_z$ dependence 
of $Z_5$ could  allow one  to check whether
$|C|\approx|D|$. At $q_z=0$,   
$Z_5$  gives   
a certain   relation between  the  s-p phases $\phi_{CA}$ and $\phi_{BD}$.

 Finally, there are other two complex numbers, each of them can be 
  determined
by measurement of two spin-correlation parameters $C_{i,j}$ and $C_{i,jk}$:
\begin{eqnarray}
  \label{z6}
Z_6=M_1M_3^*-iM_2M_4^*,\,\,
 ReZ_6=-\frac{I}{3} C_{x,xy}, \, ImZ_6=\frac{I}{2}C_{x,z},\\
  \label{z3}
Z_7=M_4M_5^*-iM_3M_6^*,\,\,
 ReZ_7=-\frac{I}{3} C_{z,yz}, \, ImZ_7=\frac{I}{2}C_{z,x}.
\end{eqnarray}
The variables  $Z_6$ and $Z_7$ are also  combinations of the 
$(p_\eta-ip_0)^{-1}$  and $(p_\eta+ip_0^*)^{-1}$ terms.
Note that all  $Z_i$ ($i=1,\dots, 7$) 
 vanish in the collinear regime ($q_x=0$).
\section{Conclusion}

  A possibility of performing a complete polarization experiment
 for the reaction $pd\to ~^3He\eta $ is studied.
 Such measurement could  give, in particular, the energy dependence of
 phases  of  s-wave transition amplitudes, providing a crucial
 test for  existence
 of the quasi-bound state in the $^3He-\eta$ system near the threshold.
 Knowledge of energy dependence of the s-wave amplitudes near the
 threshold could allow one to
 determine the signs of the real and imaginary parts of the pole point in the
 complex momentum plane. In its turn, this allows one to determine
 whether the $^3He-\eta$
 state is quasi-bound or anti-bound. 
 We derived a full set of non-zero spin  observables
 for this reaction. We show that
 in collinear kinematics the
 complete polarization experiment suggests only four measurements.
 However, one cannot separate  s- and p-wave amplitudes in collinear
 regime.  We considered two different cases beyond  collinear kinematics.
 (i) In  the case of four spin amplitudes,  four observables
  $d\sigma_0$, $A_{yy}$, $A_{xx}$ and $K_x^{x'}(p)$
 completely determine moduli of the amplitudes.
 In order to get all  four amplitudes with their relative phases, one
 needs ten observables (including spin transfer coefficients)
 and three of them can be measured roughly.
 On the contrary,  knowledge of only analyzing powers and spin-correlation
 coefficients in the initial state does not allow one to determine  moduli
 of the amplitudes independently of their phases and
 suggests  performing 14 accurate 
 measurements to get all four amplitudes. (ii) In the case of five
 independent amplitudes, measurement of spin transfer coefficients 
 also simplifies the complete polarization experiment since
 allows one to determine
 five moduli of the amplitudes in terms of five observables
 $d\sigma_0$, $A_{yy}$, $A_{xx}$, $K_x^{x'}(p)$ and $K_z^{z'}(p)$.

  Furthermore,  we show that for the general case of six
 spin amplitudes the complete polarization experiment is too 
 cumbersome  and  practically
 unrealizable. In view of this complexity, 
 we suggest to measure those
 spin observables  which allow one to single out the  pole dependence of
 the s-wave amplitudes. There  are two sets of such  observables:
 $C_{z,x}$,  $K_x^{z'}(d)$, $C_{z,yz}$,  
 $K_{yz}^{z'}$ and $C_{x,z}$, $K_z^{x'}(d)$, $C_{x,xy}$,
 $K_{xy}^{x'}$.   In general case of six independent
 amplitudes,  we show also that measurement of  two analyzing powers
 $A_{xz}$ and $A_y^d$  near the threshold  could
 provide new valuable  information
 on the $^3He-\eta$ pole.  

\section*{Acknowledgment}

 I am grateful to C.~Wilkin and  A.~Kacharava for stimulating
 discussions. I wish  to recognize the hospitality of the Institut f\"ur 
 Kernphysik of the Forschungszentrum J\"ulich, where part of this work was
 carried out. 

\section{Appendix}
\setcounter{equation}{0}
\renewcommand{\theequation}{A.\arabic{equation}}

Here we present formulae for spin observables obtained in approximation by
two  s-wave ($A$, $B$) and two  p-wave
($C$ and $D$) amplitudes. The factor $I$ in  Eqs. (\ref{sigma06})
 and (\ref{I6}) is
\begin{eqnarray}
\label{I4}
 I=|A+Cq_z|^2+|B+Dq_z|^2+ |C|^2q_x^2 +2 |D|q_x^2.
\end{eqnarray}
For  analyzing powers one has (here we preserve all six spin amplitudes)
\begin{eqnarray}
\label{axx}
A_{xx}=1-\frac{3}{I}\left[ |C|^2q_x^2+|B+Dq_z|^2\right ],\\
\label{ayy}
A_{yy}=1-\frac{3}{I}\left[ |D+E+Fq_z|^2q_x^2+|B+Dq_z-Fq_x|^2\right ],\\
\label{azz}
A_{zz}=1-\frac{3}{I}\left[ |D|^2q_x^2+|A+Cq_z|^2\right ],\\
\label{axz}
-\frac{I}{3}\,A_{xz}=q_x\left \{Re(A+Cq_z)C^*-Re(B^*+D^*q_z)D\right \},\\
\label{ayd}
 \frac{I}{2}A_y^d= Im(A+Cq_z)C^*q_x-Im(B^*+D^*q_z)Dq_x, \\
\label{ayp}
-\frac{I}{2}A_y^p=\Bigl [Im(B+Dq_z)C^*- \nonumber \\
-Im(B+Dq_z-Fq_x^2)(D^*+E^*+F^*q_z)
+Im(A+Cq_z)D^*
\Bigr ]q_x.
\end{eqnarray}
%
%spherical tensors  \cite{ohlsen}: 
%Eqs.(\ref{axx}), (\ref{ayy}),(\ref{azz}),(\ref{axz}),(\ref{ayd})
Eqs.(\ref{axx}) --(\ref{ayd})
 are equivalent to the spherical tensors  found in
 \cite{cwilkin07}:
$t_{22}=\frac{1}{2\sqrt{3}}(A_{xx}-A_{yy})$,
$t_{21}=-\frac{1}{\sqrt{3}}A_{xz}$, 
$t_{20}=A_{zz}/\sqrt{2}$, and $it_{11}=\frac{\sqrt{3}}{2}A_y^d$.

 The  proton-deuteron spin-tensor correlation parameters are
 the following:  
  \begin{eqnarray}
  \label{cyyy}
 C_{y,yy}=A_y^p-\frac{6}{I} Im (B+Dq_z)D^*q_x,\\
\label{cyyy2}
 -\frac{I}{2}C_{y,yy}=q_xIm  \{ (B+Dq_z)C^*+\nonumber \\ 
+2(B+Dq_z-Fq_x^2)(D^*+E^*+Fq_z)+
(A+Cq_z)D^* \},\\
\label{cyzz}
  C_{y,zz}=A_y^p+\frac{6}{I} Im D^*(A+C\,q_z)q_x,\\
\label{cyzz2}
-\frac{I}{2}C_{y,zz}=q_xIm\left\{ (B+Dq_z)C^*-(B+Dq_z)D^*-
2(A+Cq_z)D^*\right \},\\
\label{cyxx}
   C_{y,xx}=A_y^p+\frac{6}{I} Im C^*(B+D\,q_z)q_x,\\
\label{cyxx2}
-\frac{I}{2}C_{y,xx}=q_xIm\left\{(A+Cq_z)D^* -(B+Dq_z)D^*-2(B+Dq_z)C^*
\right \},\\
%   C_{y,xz}=3Im\ \left [CD^*q_x^2+(B+Dq_z)(A^*+C^*q_z)\right ],\\
\label{cxyz}
   -\frac{I}{3}C_{x,yz}=Im(B+Dq_z)(A^*+C^*q_z),\\
\label{czyz}
   -\frac{I}{3}C_{z,yz}= \left \{Im (A+Cq_z)D^*+Im(B+Dq_z)D^*\right \} q_x,\\
\label{cxxy}
  -\frac{I}{3}C_{x,xy}=\left \{Im (B+Dq_z)C^*+Im(B+Dq_z)D^*\right \}q_x\\
\label{czxy}
-\frac{I}{3}C_{z,xy}=Im CD^*q_x^2,\\
\label{cyxz}
\frac{I}{3}C_{y,xz}=Im CD^*q_x^2+Im(B+Dq_z)(A^*+C^*q_z),
%\label{Cijkzeros}
% C_{y,xy}=C_{y,yz}=C_{x,xz}=C_{z,xz}=0.
\end{eqnarray}

The spin correlation parameters of the  polarized proton
 and the vector polarized deuteron 
are the following:
\begin{eqnarray}
\label{cyy}
-\frac{I}{2}C_{y,y}=Re DC^*q_x^2+Re(A+Cq_z)(B^*+D^*q_z),\\
\label{cxx}
-\frac{I}{2}C_{x,x}=|D|^2q_x^2+Re(A+Cq_z)(B^*+D^*q_z),\\
\label{czz}
-\frac{I}{2}C_{z,z}=Re DC^*q_x^2+|B+Dq_z|^2,\\
\label{cxz}
\frac{I}{2}C_{x,z}=q_xRe (C-D)(B^*+D^*q_z),\\
\label{czx}
\frac{I}{2}C_{z,x}=q_x \left \{Re (A+Cq_z)D^*-Re (B+Dq_z)D^*\right \}.
 \end{eqnarray}

 The coefficients describing polarization transfer from the
 deuteron to the  $^3He$ nucleus can be written as
 \begin{eqnarray}
\label{kdhyy}
-\frac{I}{2}K_y^{y'}(d)
=Re\left \{ DC^*q_x^2+(A+Cq_z)(B^*+D^*q_z)\right \},\nonumber\\
\label{kdhxxs}
\frac{I}{2}K_x^{x'}(d)
=\left \{ |D|^2q_x^2-Re(A+Cq_z)(B^*+D^*q_z)\right \},\nonumber\\
\label{kdhzzs}
-\frac{I}{2}K_z^{z'}(d)
=Re DC^*q_x^2-|B+Dq_z|^2,\nonumber\\
\label{kdhzxs}
\frac{I}{2}K_z^{x'}(d)
=Re\left \{  C^*(B+Dq_z)+D^*(B+Dq_z)\right \},\nonumber\\
\label{kdhxzs}
\frac{I}{2}K_x^{z'}(d)
=Re \left \{ D^*(A+Cq_z)+D^*(B+Dq_z)\right \}.
%\label{kyyys}%
%-\frac{I}{3}K_{yy}^{y'}(d)
%=Im {D(B^*+D^*q_z)}.
\end{eqnarray}

The coefficients of polarization transfer from
 the initial proton   to the  $^3He$ nucleus are 
 \begin{eqnarray}
\label{kphyy}
I\,K_y^{y'}(p) =|C|^2q_x^2+|A+Cq_z|^2,\nonumber \\
\label{kphxx}
I\,K_x^{x'}(p)=|C|^2q_x^2+|A+Cq_z|^2 -2|D|^2q_x^2,\nonumber\\
\label{kphzz}
I\,K_z^{z'}(p)=|C|^2q_x^2+|A+Cq_z|^2 -2|B+Dq_z|^2q_x^2,\nonumber\\
\label{kphxz}
\frac{I}{2}\,K_x^{z'}(p)
=q_x Re \left \{(B+Dq_z)C^*-(A+Cq_z)D^* -(B+Dq_z)D^*\right \},\nonumber\\
\label{kphzx}
\frac{I}{2}\,K_z^{x'}(p)
=q_x Re \left \{(A+Cq_z)D^*- (C^*+D^*)(B+Dq_z)\right \}.
\end{eqnarray}
% Eqs. (\ref{axx}),  (\ref{axz}), (\ref{ayd}), (\ref{ayp}),
 Eqs. (\ref{axx}) -- (\ref{ayp}),
 (\ref{cyyy2}),
  (\ref{cyy})  are valid 
 in the general case of six spin amplitudes.

\end{document}